\documentclass[%
 aapm,
 mph,%
 amsmath,amssymb,longbibliography,
 reprint,%
]{revtex4-1}

\usepackage{graphicx}
\usepackage{dcolumn}
\usepackage{bm}
\usepackage{amsmath,amssymb}

\usepackage{hyperref}
\usepackage[]{natbib}

\usepackage{tikz}
\usepackage{lipsum}

\usepackage[mathlines]{lineno}

\begin{document}


\title{Threshold in quantum correlated interference for a particle interacting with two scatterers}


\author{F. V. Kowalski}
 \affiliation{Physics Department, Colorado School Mines.}
 \email{fkowalsk@mines.edu.}

\date{\today}

\begin{abstract}
Correlated interference is calculated for a microscopic particle retro-reflecting from two spatially separated scatterers that are free to move, all three of which are treated as quantum bodies: the positions of the particle traversing this one-dimensional interferometer and those of the interferometer components are all uncertain. Interference in a measurement of only the particle is shown to disappear for microscopic yet appear for mesoscopic scatterers, contrary to that of a quantum-classical transition. A measurement of this threshold can verify quantum correlations in many-body systems by measuring only the retro-reflected microscopic particle. The decoherence of a mesoscopic scatterer is determined by this threshold without it having to traverse slits or beamsplitters.
\end{abstract}

\maketitle

\section{\label{sec:intro}Introduction}


Quintessential illustrations of non-classical phenomena are found in many-body quantum correlations. EPR experiments seem to be non-interferometric examples \cite{reid} although they have been interpreted as such \cite{horne}. A straightforward example of quantum correlated interference, QCI, is given by Gottfried \cite{gottfried}.

The vast majority of QCI studies in the literature involve bipartate photon states traversing interferometers whose components are treated as rigid classical potentials \cite{kim}. The limitations on measurement precision due to the quantum nature of an interferometer component has been discussed \cite{braginsky} but not related to QCI. Micromechanical oscillators as components of interferometers have been treated as quantum objects \cite{aspelmeyer}, but again not related to QCI. QCI between the interferometer components and the particle traversing it is considered less often \cite{tomkovic,chapman,kowalski1,kowalski2}. 


A many-body treatment of a particle Bragg scattering from a crystal that is comprised of uncoupled atoms in a ordered lattice is an example of correlated interference between the particle and interferometer components (which are in this case the scatterers). The initial approach to understanding such scattering is to use the same approximation as with bipartate photon states traversing an interferometer: the scatterers are rigid classical potentials (and therefore essentially localized) while only the particle is treated as a quantum object. However, the issue of concern below is the effect of uncertainty in the scatterer positions on interference. While it is not practical to vary the uncertainty of the atom positions in a real crystal it may be feasible after the atoms in an optical lattice are released from their confining potential \cite{bloch}. 

In a crystal of mass $M=Nm$, where $N$ is the number of constituents each of mass $m$, the uncertainty in each atom's position results in an uncertainty in the center of mass, c.m., of the crystal. For an object in thermal equilibrium with the environment, the coherence length of its c.m. is
$L_{c}^{thermal} \approx \lambda^{2}/\Delta \lambda = \lambda V/\Delta V$ where $\lambda$ is its thermal deBroglie wavelength, $V$ is its velocity \cite{hasselback} and $\Delta V_{thermal} \approx \sqrt{2 k_{B} T/M}$, yielding $L_{c}^{thermal} \approx h/\sqrt{2Mk_{B}T}$. This dependence of $L_{c}^{thermal}$ on $1/\sqrt{Nm}$ implies that the uncertainties in the individual atom positions (presumably of value $h/\sqrt{2mk_{B}T}$) add in quadrature to yield the uncertainty in the mean position of the c.m. of the crystal. If the crystal is a component of an interferometer, such as a mirror, then its coherence length affects correlated interference between the particle and c.m. of the mirror (a component of the interferometer), as discussed in the next section. 

One consequence of QCI, described below, is that interference when measuring only the particle disappears as the uncertainty in the scatterer position becomes comparable to the particle wavelength. Such a transition in interference is then expected to also occur in other interferometers (for example, in Bragg scattering).

A related elimination of interference in the one particle marginal distribution has been described in the context of QCI for two particles, both generated in a single momentum-conserving decay (momentum conservation also generates correlation in the particle-scatterers system described here), while each traverses a different ``classical'' double slit potential \cite{gottfried}. A corresponding experiment confirmed such elimination of marginal interference \cite{waitz}. Calculations on a pair of correlated particles that traverse interferometers whose components are treated as classical potentials also predict marginal distributions that vanish \cite{horne,peled}. Two-body interference has been observed with a Bose-Einstein condensate that uses a potential well as a classical double slit but marginal distributions were not measured \cite{borselli}. Marginal distributions also play an important role in quantum computing due to the difficultly in measuring all of the qubits after completion of a computation \cite{jacobs}.



Classical interferometric transitions occur as the polarizations of the superposed waves become orthogonal or as the coherence length becomes smaller than the path differences in an interferometer. The spatial interference pattern also vanishes as the particle wavelength becomes (a) larger than the slit spacing in double slit interferometer and (b) larger than the scatterer spacing in Bragg interference \cite{Note1}. In quantum systems, an addition transition, described by decoherence theories \cite{schlosshauer} or decoherent histories \cite{griffiths}, occurs due to quantum correlations between a body in a superposed state and the many degrees of freedom associated with the environment to which it is coupled. Yet another transition occurs as a consequence of a breakdown in the Schr\"odinger equation, models of which predict a quantum-classical transition \cite{leggett2002}. 

Contrary to the results expected in such quantum-classical transitions, the transition discussed below involves interference that is detectable for mesoscopic but not microscopic scatterers. Although this occurs for an isolated system with few degrees of freedom, it is also the result of quantum correlations (between the particle and scatterers rather than with the environment). 

These issues are first introduced using the much simpler two-body quantum system of a particle elastically retro-reflecting from a ``mirror'' whose c.m. position is uncertain (forming two-body ``standing wave'' interference) \cite{kowalski1,kowalski2}. However, it is difficult to devise a practical apparatus using such a two-body system that facilitates measurement of only the particle to verify the interferometric transition. To address this a particle-two-scatterer model is introduced and possible methods for measuring the transition are discussed for both micro and mesoscopic scatterers.


\section{\label{sec:particlemirror}Particle-mirror model}

\begin{center}
\begin{figure}
\includegraphics[scale=0.29]{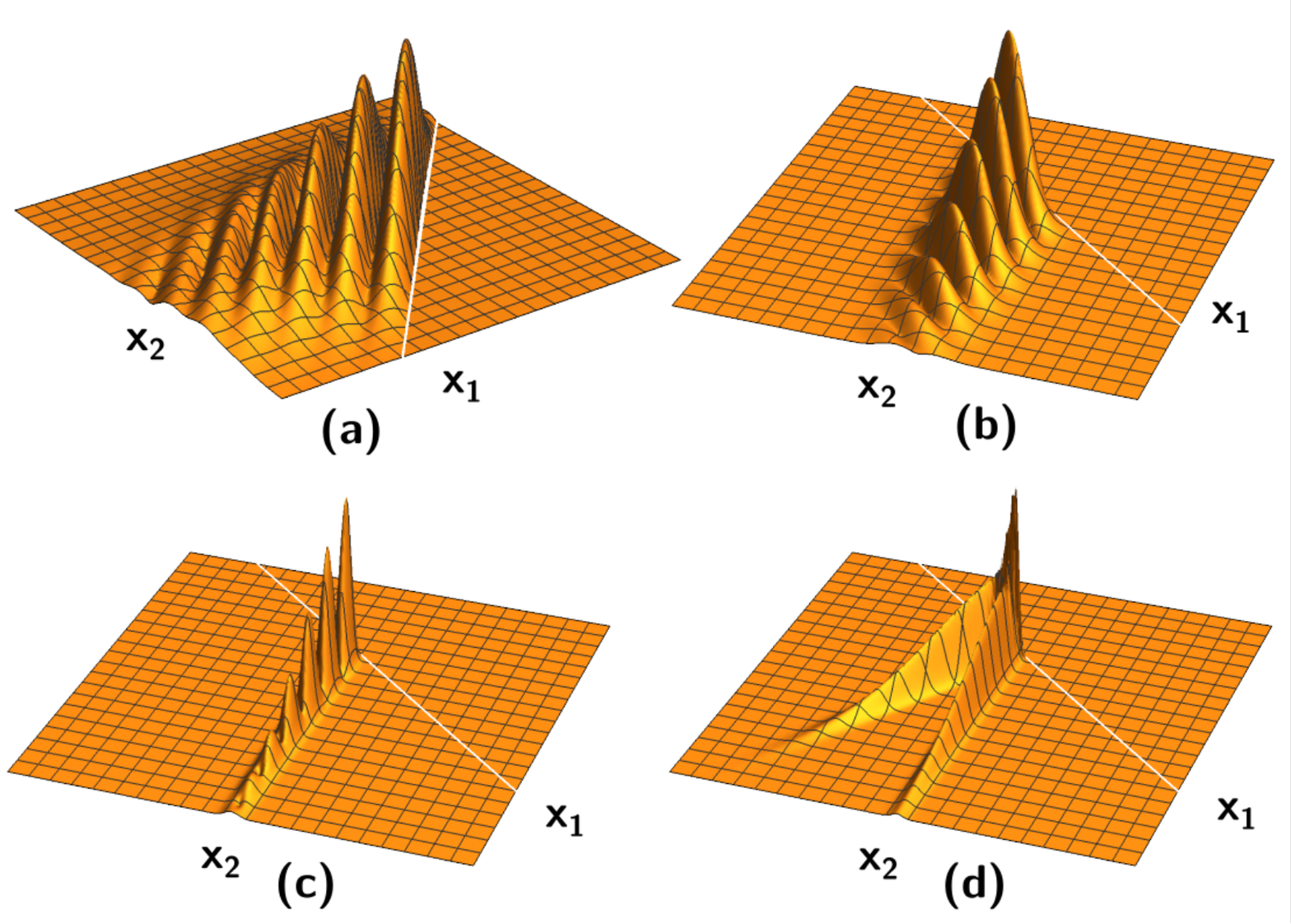}
\caption{Two-body PDF plots for a particle retro-reflecting from a mirror. The mirror coherence length changes sequentially between the panels while the particle coherence length is fixed. The distance spanned along both axes is the same in all panels. In panels (a), (b), and (c) $M/m=100$ while in panel (d) $M/m=5$ revealing the effect of recoil on a mirror with a small coherence length.}
\label{fig:bandwidth}
\end{figure}
\end{center}

A particle, with mass $m$, coordinate location $x_{1}$ and speed $v$, retro-reflects from the center of mass, c.m., of a scatterer or ``mirror'' with mass $M$, location $x_{2}$, and speed $V$ where $m\ll M$ and $v\ll V$. The PDF shown in fig. \ref{fig:bandwidth} illustrates QCI between incident and reflected two-body wavegroups (related figs. are found in \cite{kowalski1,kowalski2}). The coherence length of the mirror becomes progressively smaller between the panels shown in this figure. 

A position measurement of only the particle that reveals interference for the probability density function, PDF, shown in fig. \ref{fig:bandwidth}(a) requires a spatial resolution smaller than the peak-to-peak variations in the PDF, which are of order the particle wavelength. To calculate the marginal probability density function, the center of mass, c.m., degree of freedom of the mirror is traced out of the pure state density matrix, which involves an integration of the PDF over the mirror coordinate. The interferometric oscillations then vanish for fig. \ref{fig:bandwidth}(a) but appear when the coherence length of the mirror, $L_{c}$, is reduced as shown in panels (b) and (c). The interferometric transition in the particle marginal PDF, the focus of this work, therefore occurs between fig. \ref{fig:bandwidth}(a) and (c) panels. 

However, as the mirror's $L_{c}$ increases (often as its mass decreases), interference in the mirror's substate appears. Marginal particle interference then disappears, operationally defining $L_{c}$. In such a determination of $L_{c}$, the mirror is not required to traverse division of amplitude or division of wavefront interferometers to determine its quantum behavior. 

On the other hand, the marginal distribution for the mirror involves an integration of the PDF shown in fig. \ref{fig:bandwidth} over the particle coordinate. There is then no transition comparable to that of the particle marginal distribution in a measurement of only the mirror for the particle substate coherence length shown. Classical reflection of an electromagnetic wave from a mirror similarly results in only interference of the wave but not the mirror. While the quantum and classical predictions are consistent in this case of measuring only one body they deviate in a correlation measurement of both the particle and mirror. For example, fig. \ref{fig:bandwidth} (c) illustrates the classical analog of the quantum result of reflection from a mirror (its meso-macroscopic mass has a small $L_{c}^{thermal}$). Correlated interference nevertheless exists: both the particle and the mirror will not be found at certain locations in the $x_{1},x_{2}$ plane. Classically there is no position where the mirror is not found.

Two-body PDF plots for a particle retro-reflecting from a mirror. The mirror coherence length changes sequentially between the panels while the particle coherence length is fixed. The distance spanned along both axes is the same in all panels. In panels (a), (b), and (c) $M/m=100$ while in panel (d) $M/m=5$ revealing the effect of recoil on a mirror with a small coherence length.

Interference is a consequence of not knowing if the particle reflected from the mirror or not. Particle and mirror Gaussian substates with their full width half maximum, FWHM, values much smaller (not shown in  fig. \ref{fig:bandwidth}) than the interferometric oscillations in the PDF essentially exhibit reflection behavior of classical point bodies. No interference occurs since a position measurement of either the particle or mirror or both reveals that the particle either reflected and the mirror recoiled or not.

\section{\label{sec:particlescatterers}Particle-two-scatterer model}

The constraints of high spatial resolution and segregated particle-mirror measurement are difficult to satisfy in any practical attempt to measure the transition associated with the particle marginal PDF in fig. \ref{fig:bandwidth}. This is mitigated with a three-body interferometer. A particle of mass $m$, initial speed $v$, and coordinate location $x_{1}$ along the x-axis retro-reflects (s-wave scattering in the Born approximation) from the scatterers that also move along the x-axis. They have mass $M_{2}$ and $M_{3}$, initial speeds $V_{2}$ and $V_{3}$, and coordinate locations $x_{2}$ and $x_{3}$, respectively (a one scatterer version is shown schematically in fig. \ref{fig:gaussian}).

The PDF for these three uncoupled bodies is derived from two scattering amplitudes, each of which is a three-body wavefunction. There is an amplitude for the  particle to retro-reflect from only mass $M_{2}$, which is initially located at the origin and there is an amplitude for it to similarly retro-reflect from mass $M_{3}$, which is initially a distance $x_{0}$ away from the origin. These two three-body scattered amplitudes are:  $\psi_{12}^{scattered}\psi_{3}$ and $\psi_{13}^{scattered}\psi_{2}$ where the numerical subscripts refer to each body in the same manner as with the coordinates. Weak scattering is assumed to minimize the effect of multiple reflections.

Consider interference of only the three-body scattered amplitudes. The incident three-body state then has a negligible effect on the PDF at the observation point, thereby eliminating interference between the incident and two scattered amplitudes (unlike that shown in fig. \ref{fig:bandwidth}), similar to the interference of an EM wavegroup observed in reflection far from a ``thin-film'' interferometer. To accomplish this a particle wavegroup substate is required to have a coherence length much greater than the scatterer separation but smaller than the distance between the observation point and the scatterers. The three-body amplitude for the particle to scatter from $M_{3}$ is then a product of the initial $M_{2}$ scatterer substate, $\psi_{2}$, with the entangled scattered substate $\psi_{13}^{scattered}$. A related result follows for the scatttering from $M_{2}$. Details are found in the appendix \ref{appendix:eigenstates}.

For mesoscopic scatterers and a microscopic particle let $M_{2}, M_{3} \gg m$ and $v\gg V_{2}, V_{3}\approx 0$, yielding the three-body joint PDF for eigenstates of momentum
\begin{equation}
\begin{gathered}
\textrm{PDF}^{M \gg m}[x_{1},x_{2},x_{3}] \approx \cos^{2}[\frac{mv(x_{3}-x_{2})}{\hbar}].
\label{eqn:main}
\end{gathered}
\end{equation}
The coordinates within the brackets in the notation $\textrm{PDF}^{M \gg m}[x_{1},x_{2},x_{3}]$ indicate correlated interference between the bodies associated with each coordinate even if, as in this case, the PDF itself does not explicitly contain all of these coordinates. A substate interferometer is classified as ``closed'' if the PDF does not depend on its coordinates. For example, interference in a measurement of only the particle does not depend on its coordinate in eqn. (\ref{eqn:main}) (this is also the case for a particle traversing a Michelson interferometer). Otherwise  it is``open'' as is illustrated in  eqn. (\ref{eqn:main}) for a measurement of either $M_{2}$ or $M_{3}$ (this is also the case for a particle traversing a double slit) \cite{kowalski2}.

To incorporate the effect of wavegroups into the interference predicted by eqn. (\ref{eqn:main}) let the particle have a Gaussian substate with a coherence length much larger than $2x_{0}$ while the scatterers have  Gaussian substates with coherence lengths less than $x_{0}$. The particle and scatterers will typically be observed only within the spatial regions defined by these wavegroups. 

Let $x_{3} = \overline{x_{3}}+ \delta x_{3}$, where $\overline{x_{3}}\approx x_{0}$ is the location of the Gaussian maximum of $M_{3}$ and where $\delta x_{3}$ is a distance that varies only of order a few FWHMs from this peak (beyond which the PDF essentially vanishes). Corresponding parameters describe the other scatterer resulting in the three-body PDF
\begin{equation}
\textrm{PDF}^{M \gg m}[x_{1},x_{2},x_{3}] \propto \cos^{2}[\frac{mv(x_{0}+\delta x_{3}-\delta x_{2})}{\hbar}]
\label{eqn:correlation}
\end{equation}

If the scatterer's Gaussian widths satisfy $mv \delta x_{2}/\hbar \ll 2\pi$ and $mv \delta x_{3}/\hbar \ll 2\pi$, then
\begin{equation}
\begin{gathered}
\textrm{PDF}^{M \gg m}[x_{1},x_{2},x_{3}]\propto \cos^{2}[mvx_{0}/\hbar].
\label{eqn:correlation0}
\end{gathered}
\end{equation}
The particle marginal PDF then yields interference mathematically similar to that from two ``classical'' scatterers. 

If the scatterer substates are so large that it is likely for the scatterers to be found in locations away from their peak values then $mv \delta x_{2} > 2\pi \hbar$ and $mv \delta x_{3} > 2\pi \hbar$. This results in eqn. (\ref{eqn:correlation})
which describes three body quantum correlated interference. Interference maxima are associated with locations at which the particle and scatterers are most likely to be found: a particle will be found somewhere in retro-reflection when the scatterers are found in positions that satisfy $mv(x_{0}+\delta x_{3}-\delta x_{2})=\pi n \hbar$, with $n$ an integer.

A plot of eqn. (\ref{eqn:correlation}) appears similar to that shown in fig. \ref{fig:bandwidth}(a) with the substitution $x_{1} \rightarrow \delta x_{3}$ and $x_{2} \rightarrow \delta x_{2}$ for the axes in this figure. The fringe spacing along the $\delta x_{3}$ axis, from eqn. (\ref{eqn:correlation}), is $\approx h/2m v$ for fixed $\delta x_{2}$. The particle marginal interference essentially disappears for
\begin{equation}
\begin{gathered}
L_{c} >  \lambda_{0}/2.
\label{eqn:transition}
\end{gathered}
\end{equation}
Note that this condition requires an open interferometer for at least one of the scatterers for the integration along that scatterer coordinate to eliminate interferometric oscillations.

The particle that retro-reflects from the scatterer can be a photon. Fedorov's example of a correlated atom-photon two-body wavefunction \cite{fedorov} results in replacing the wavelength in the transition condition, eqn. (\ref{eqn:transition}), with that of the photon.  

An extension of eqn. (\ref{eqn:correlation}) to three identical scatterers yields the four-body PDF
\begin{equation}
\textrm{PDF}^{M \gg m} \propto 3+2(\cos[a_{23}] +\cos[a_{24}]+\cos[a_{34}]),
\label{eqn:fourscatterers}
\end{equation}
where $a_{jk}=mv(x_{jk}+\delta x_{k}-\delta x_{j})/\hbar$, $x_{jk}=x_{0j}-x_{0k}$, and $x_{0i}$ is the initial position of the $i^{th}$ scatterer for $i,j,k$ going from $2$ to $4$. This begins to approximate one dimensional Bragg scattering. Note that the visibility of the particle marginal interference is reduced if only one scatterer satisfies  $mv \delta x_{j}/\hbar > 2\pi$ since the corresponding term in eqn. (\ref{eqn:fourscatterers}) vanishes in an integration over that scattterers coordinate (as it is traced out of the density matrix). Particle marginal interference is eliminated if only two of the three scatterers satisfy this constraint.

If the particle is treated as a quantum body while the interferometer components are classical potentials then elimination of interference is typically associated with interferometer path differences that are greater than the coherence length of the particle. This also occurs in many-body interference if the coherence length of the particle substate is less than $2x_{0}$. 

However, an additional constraint is required for interference in a many-body system: recoil in the interaction must not be sufficient to spatially separate the scatterer substates of having and not having reflected the particle. If these scatterer states no longer overlap then interference is not manifest in any measurement of the many-body system. This is illustrated in fig. \ref{fig:bandwidth}(d) where the recoil of the mirror is sufficient to separate the mirror substates of having and not having reflected the particle \cite{Note2}. Such momentum exchange is assume here to be negligible.

\section{\label{sec:particlescatterer}Particle-one-scatterer model}

Next replace the scatterer at the origin with a macroscopic beamsplitter, bs, which is not measured. To do so let $m\delta x_{2}/\hbar \ll 2\pi$. The bs and the $M_{3}$ scatterer are assumed to have the same small scattering amplitude resulting in (see appendix \ref{appendix:wavegoups})
\begin{equation}
\begin{gathered}
\scalebox{1.}{${\displaystyle \textrm{PDF}^{\textrm{bs}}[x_{1},x_{3}] \propto  {\bf A} \cos^{2}[\frac{2\pi x_{3}}{\lambda_{0}}]}$},\\
{\bf A}\propto {\rm Exp}[-\frac{8 \pi^{2}(x_{3}-x_{0})^{2}}{L_{c}^{2}}]/\sqrt{ L_{c}}.
\label{eqn:correlation2}
\end{gathered}
\end{equation}

\begin{center}
\begin{figure}
\includegraphics[scale=0.28]{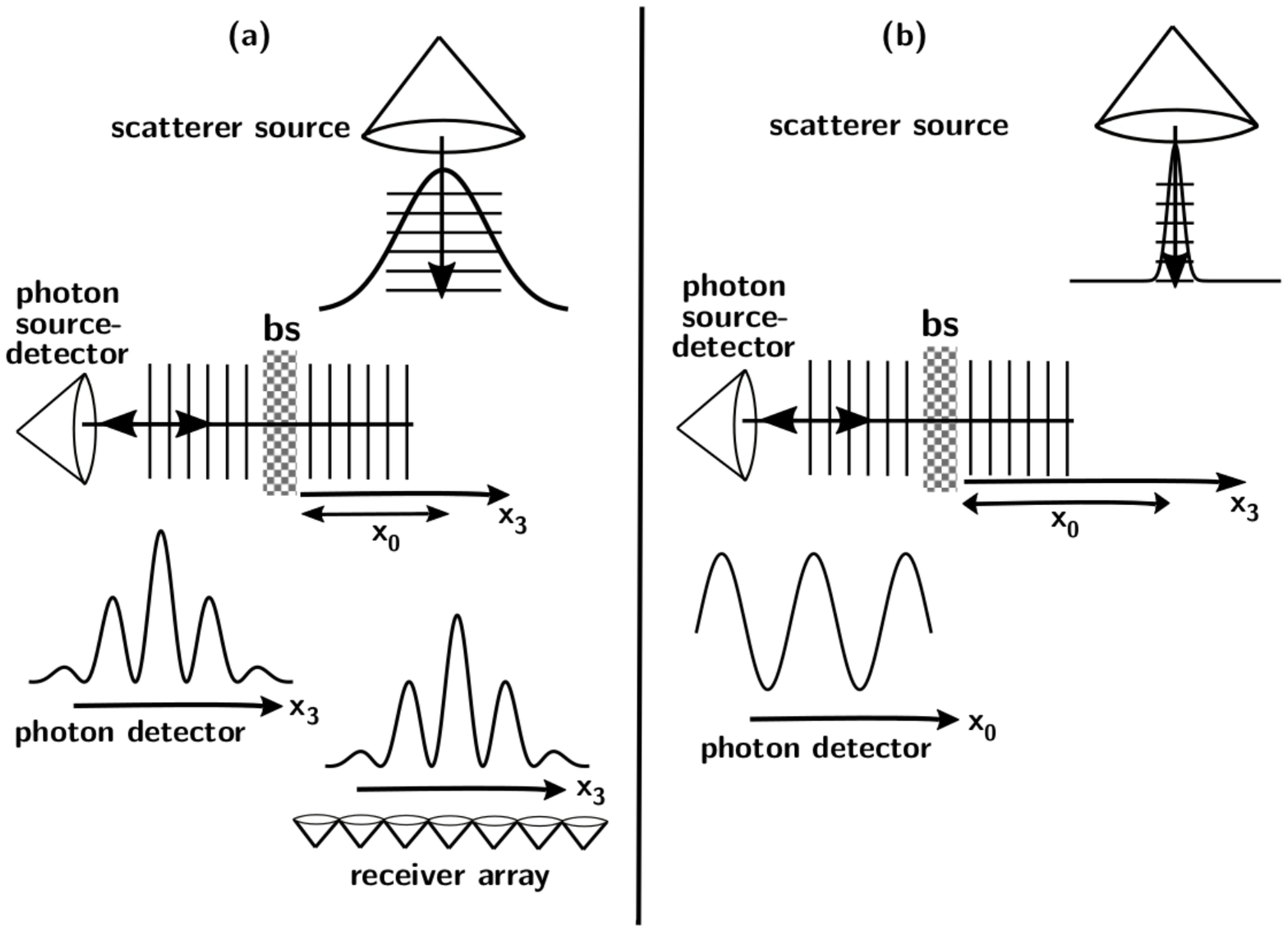}
\caption{Particle-one-scatterer apparatus where bs is the beamsplitter. Correlated interference measured with a photon detector and a scatterer receiver array is shown in panel (a) where $x_{0}$ is fixed (the interference is similar to that shown in fig. \ref{fig:bandwidth} (a)). The photon marginal distribution in this case yields no interferometric oscillations. Panel (b) illustrates a photon marginal distribution for a short transverse coherence length scatterer with variable position $x_{0}$. The interferometric transition in the photon marginal distribution occurs between (a) and (b).}
\label{fig:gaussian}
\end{figure}
\end{center}

Let a beam of $M_{3}$ scatterers move perpendicular to the particle beam as shown in fig. \ref{fig:gaussian}. The center of the scatterer beam intersects with the particle path at position $x_{0}$ while its FWHM distance is projected along the particle trajectory with a Gaussian profile. The motion of the scatterer perpendicular to the particle facilitates the measurement of scatterer's interferometric oscillations as a function of $x_{3}$ with a detector array. Assume that the beam density allows only one particle to interact with one scatterer in the beam on average. 

To illustrate a method to measure both correlated and marginal interference let the scatterers be atoms in the beam with a transverse coherence length $L_{c}\propto \textrm{FWHM}$ while the particle is a photon. The correlated interference described in eqn. \ref{eqn:correlation2} with $x_{0}$ fixed is shown schematically in panel (a) of fig. \ref{fig:gaussian}. If the photon is measured to retro-reflect then the scatterer is measured at a random position along the $x_{3}$ axis (within the scatterer Gaussian envelope). Multiple such events recorded in aggregate reveal the photon-scatterer correlated interference given by eqn. (\ref{eqn:correlation2}). A minimum corresponds to no observation of the photon in retro-reflection and none of the scatterer at certain locations along the receiver array (or $x_{3}$). If the photon is measured to transmit then the random distribution of scatterer measurements along the $x_{3}$ axis exhibits a Gaussian distribution with no interferometric oscillations.

Results from measurements of only the photon are predicted by
\begin{equation}
\begin{gathered}
\textrm{PDF}^{\textrm{bs}}[x_{1}] =\int_{\infty}^{\infty} \textrm{PDF}^{M \gg m}[x_{1},x_{3}] dx_{3}\\
\approx 1-{\rm Exp}[-\frac{L_{c}^{2}}{2\lambda_{0}^{2}}] \cos[{\displaystyle \frac{4\pi x_{0}}{\lambda_{0}}}].
\label{eqn:marginalPDF}
\end{gathered}
\end{equation}

Eqn. (\ref{eqn:correlation0}) is a consequence of eqn. (\ref{eqn:marginalPDF}) for $L_{c}\ll \lambda_{0}$. However, marginal interference essentially vanishes for $L_{c}> \lambda_{0}$, which agrees with the result given in eqn. (\ref{eqn:transition}). As the scatterer coherence length becomes less than $\lambda_{0}$ the photon marginal distribution (given in eqn. (\ref{eqn:marginalPDF})) exhibits interference as a function of $x_{0}$ as shown in panel (b) of fig. \ref{fig:gaussian}. The transition in photon marginal interference, from its elimination for $L_{c}> \lambda_{0}$ to its existence for $L_{c} \leqslant \lambda_{0}$ then operationally defines the scatterer coherence length via eqn. (\ref{eqn:marginalPDF}).


As the scatterer's transverse coherence length decreases, its substate probability distribution is essentially non-zero only for $x_{3}=x_{0}$. Varying $x_{0}$ in eqn. (\ref{eqn:correlation2}) then yields correlated interference between both the photon and scatterer. This models QCI for meso-macroscopic scatterers with small thermal coherence lengths that occurs in fig. \ref{fig:gaussian} (b) and and fig. \ref{fig:bandwidth}(c). Although it is not shown explicitly in fig. \ref{fig:gaussian} (b) a null in the measurement of the photon correlates with no measurement of the scatterer anywhere along the $x_{3}$ axis (the photon transmitted rather than reflected). If, on the other hand, no interferometric correlation existed between these measurements of the photon and scatterer then a maximum in a measurement of the photon distribution could occur for a null in a measurement of the scatterer distribution: the photon reflected without scatterer recoil. Correlated interference is required to prevent such violation of conservation laws. 

However, if this superposition state of the scatterer in fig. \ref{fig:gaussian} (b) (or the mirror in fig. \ref{fig:bandwidth} (c)) decoheres due to coupling with the environment then interferometric oscillations in the photon (or particle) marginal distributions also disappear: the scatterer (or mirror) is no longer in a superposition of having and not having reflected the particle. In many-body interference a superposition of substates for each body of having and not having reflected must exist. In addition, the substates for each body must overlap. 


If coupling of the scatterer to the environment eliminates scatterer superposition then a commensurate lack of marginal interference provides a method to investigate meso-macroscopic scatterer decoherence mechanisms without measuring the scatterer. On the other hand, if scatterer coupling to the environment only reduces the coherence length of the scatterer while the scatterer remains in a superposition state then both marginal and correlated interference remain (as long as recoil does not sufficiently separate the superposed scatterer substates to prevent overlap). 




\begin{center}
\begin{figure}
\includegraphics[scale=.9]{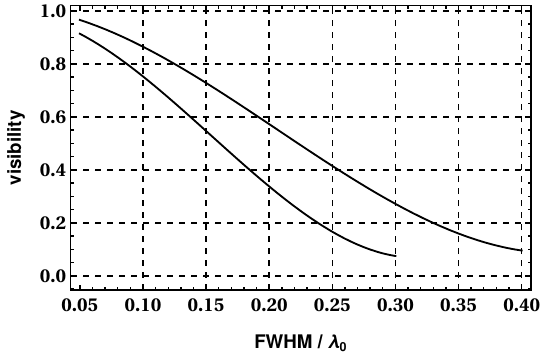}
\caption{The visibility of the particle marginal distribution is determined by measuring the interferometric oscillations for the apparatus shown in fig. \ref{fig:gaussian} (b) while varying $x_{0}$. The upper trace is the result for the particle-one-scatterer model while for the lower trace the beamsplitter is replaced with a scatterer identical to that used in the upper trace. Both traces are shown as a function of the ratio of the FWHM of the scatterer's substate Gaussian PDF (for the scatterer before interaction) divided by the incident particle wavelength. The values used are $m/M=1/1200$ with the particle coherence length greater than $x_{0}$ by a factor of $30$ while the particle wavelength is less than $x_{0}$. The ratio of the scatterer FWHM to $x_{0}$ varied approximately from $0.01 \rightarrow 0.5$.}
\label{fig:visibility}
\end{figure}
\end{center}

The visibility of the particle marginal distribution for the particle-one-scatterer model, shown as the upper trace of fig. \ref{fig:visibility}, is given as a function of the ratio of the single scatterer FWHM to the fixed incident particle wavelength. The FWHM of a Gaussian wavegroup is proportional to its coherence length as is the visibility function \cite{born}. This then is essentially a plot of the coherence length of the particle in a marginal measurement as a function of the coherence length of the scatterer. Classically, visibility characterizes a property of the wave traversing an interferometer in which there are no uncertainties in the positions of the components comprising the interferometer.

The results shown in fig. \ref{fig:visibility}, however, differ from this standard interpretation. Since the coherence length of the particle is much larger than that of both the scatterer and of $x_{0}$, the visibility function shown in this figure corresponds to that of the scatterer, not the particle (although only the particle is measured). It reveals information about the many-body system and does not, as in the classical case, exclusively characterize properties of the wavegroup (particle) traversing the interferometer. The upper trace demonstrates the effect of uncertainty in one interferometer component (the scatterer) on particle marginal interference. It again confirms the result that the particle marginal distribution exhibits interference when the scatterer coherence length is less than the particle wavelength and eliminates this interference in the converse case.

The lower trace corresponds to the particle marginal visibility function in the particle-two-scatterer model. Fig. \ref{fig:gaussian} (b) is the result of replacing the beamsplitter with an identical scatterer while their coherence lengths are simultaneously varied by the same amount. Although not shown in figure 3, the visibility function for correlated interference in the particle-two-scatterer model when measuring all three bodies yields a value of $0.8$ for FWHM/$\lambda_{0}=0.3$. The visibility results shown for both the particle-one-scatterer and particle-two-scatterer models are derived from the three-body states described in the appendix.

Consider relating the two visibility functions shown in fig. \ref{fig:visibility} to $L_{c}^{thermal}$, described in the introduction, by allowing both scatteres to be ``reflectors'' of equal mass. The superposition of two harmonic waves (those reflected from each scatterer) of equal amplitude can be expressed as a single harmonic wave whose amplitude and phase are determined by $x_{0}$. Assume that such interference in retro-reflection is similar to reflection from only one scatterer of mass $2M$. A possible justification is that the scatterers are entangled via conservation of energy and momentum in reflection of the particle. A measurement of one then affects both scatterers. Therefore, rather than recoiling from an individual scatterer, the particle could in effect scatter from the mass of the entangled scatterer system \cite{kowalskicollective}. $L_{c}^{thermal}$, proportional to the visibility function, for this artificial scatterer of mass $2M$ then decreases by a factor of $1/\sqrt{2}$ from that of one scatterer of mass $M$. This is apparent in  fig. \ref{fig:visibility} by choosing a point on the upper trace and transforming its horizontal component by a factor of $1/\sqrt{2}$, thereby reducing its coherence length, which is proportional to $1/\sqrt{M}$, to one proportional to $1/\sqrt{2M}$. The transformed point then falls on the lower trace.


\section{\label{sec:discussion} Discussion}


The expression for $L_{c}^{thermal}$ yields, in the particle-one-scatterer model, the following constraint on the scatterer mass for particle marginal interference to disappear,
\begin{equation}
\begin{gathered}
M_{3} < \frac{2 h^{2}}{\lambda_{0}^{2}  k_{B} T},
\label{eqn:massIneq}
\end{gathered}
\end{equation}
assuming that the component of the scatterer velocity along the particle direction is essentially zero. As a simple illustration of the boundary let the particle be a neutron of mass $m_{n}$ retro-reflecting from a scatterer which has $M_{3}>m_{n}$. The boundary occurs at $M_{bndry} \approx 23000 ~m_{n}$, for $T=1^{\circ}$K and $v_{n}=10^{4}$m/s ($\lambda_{0} \approx 0.04$ nm).

However, the boundary of the interferometric transition is not fundamentally limited by the thermal coherence lengths of the scatterers if the scatterers can be isolated from the environment. Each scatterer can be prepared in a state with a coherence length longer than $L_{c}^{thermal}$. When exposed to the environment it decoheres to $L_{c}^{thermal}$ with an environmental relaxation time constant \cite{corbitt,miao,matsumoto2016}. Measurements made on a time scale shorter than this relaxation time can in principle reveal quantum behavior even at room temperature and for macroscopic masses (see the discussion of eqn. $1.30$ in ref. \cite{braginsky}).


An experimental result related to the particle-one-scatterer model involves a photon as the particle and an atom and macroscopic mirror as the scatterers \cite{tomkovic}. This is analogous to the following modifications of the apparatus described in sec. \ref{sec:particlescatterer}: (1) the functions of the bs and atom scatterer and their locations are interchanged with the bs becoming a mirror, (2) the atom recoil momentum is sufficient to spatially separate the atom states of having and not having retro-reflected the photon (thereby eliminating interference), similar to the mirror substates shown in fig. \ref{fig:bandwidth}(d), (3) these atom states are recombined with a grating downstream from the interaction region (where the atom substates now overlap, generating correlated interference). Since the particle and atom substate interferometers are now closed there is no particle marginal interferometric threshold. Yet QCI is manifest in the atom marginal interference that varies with displacement of the mirror (with which only the photon interacts). Related calculations of QCI on systems with closed substate interferometers are found in \cite{kowalski2}. 


The marginal interferometric transition has characteristics of a decoherence mechanism; the elimination of interference is accomplished by the neglect of quantum correlation information. In spite of there being no interference in the measurement of only the particle, the system remains in a superposition state: interference is resurrected if all the bodies are measured. 

Yet differences between the marginal and environmental decoherence transitions exist. Decoherence theory is focused on explaining why macroscopic superposition states are not observed due to the large number of degrees of freedom associated with environmental interactions rather than on the elimination of interference in a marginal distribution for a microscopic body interacting with a meso-macroscopic body as described above. 

The results presented here indicate an additional difference. Meso-macroscopic mirror correlated interference in the particle-mirror model, as shown in fig. \ref{fig:bandwidth} (c) (or particle scatterer correlated interference in the particle-one-scatterer model), could in principle be observed due to the small number of degrees of freedom in this system. However, such a measurement requires an apparatus with an appropriate interaction Hamiltonian between the center of mass of the scatterer and a pointer state of the measuring device, a difficult proposition for macroscopic bodies. 

Without verification of such correlated interference a measurement of only the long coherence length particle or only the meso-macroscopic mirror in the particle-mirror system yields a result consistent with the classical interpretation: interference is observed for only the long coherence length ``wave'' (particle) but not the mirror. This is also the case for measurement of only the particle or only the meso-macroscopic scatterer in the particle-one-scatterer model. Observation of only particle marginal interference (as is done in the most interferometric measurements) is therefore consistent with either the interpretation that the unmeasured superposed meso-macroscopic interferometer components have not decohered or that the interferometer has passed from the quantum to classical domain where many-body interference (and therefore the marginal transition) does not exist. 

However, these interpretations are distinguishable if the transition in particle marginal interference is observed: the mirror must then be in a superposition state even if the mirror is not observed. This non-classical effect occurs for the particle-one-scatterer system when the scatterer (with its $L_{c} > \lambda_{0}$), the macroscopic bs (with its $L_{c} \ll \lambda_{0}$), and the particle (with its $L_{c} > x_{0}$) are all in a superposition state of having and not having reflected the particle. It can also occur if the mirror or scatterer, when exposed to the environment, decoheres, eliminating correlated many-body interference and therefore also interference in the particle marginal distribution. 


The above issues are a consequence of three cornerstones of quantum physics: (a) the superposition principle that allows both the particle and scatterer to be in states of having and not having interacted, (b) the large coherence lengths possible for the particle and scatterer, a result of the uncertainty principle, and (c) measurements of marginal distributions, a consequence of the Born rule as applied to discarding information in QCI. 

Experimental verification of this interferometric transition (requiring at least one substate open interferometer) will yield insight into quantum correlations, the coherence lengths of mesoscopic bodies, decoherence, and the effects of measurement, or lack thereof, on many-body systems without having to measure all of the bodies in a many-body system.

\appendix
\section{\label{appendix:eigenstates}{\bf Three-body eigenstates of momentum}}

Assume that before interaction the two scatterers and particle are initially in separable eigenstates of momentum with the three-body state given by $\psi_{1} \psi_{2}\psi_{3}$, where $\hbar K_{2}=M_{2} V_{2}$, $\hbar K_{3}=M_{3} V_{3}$, and $\hbar k=m v$. The superposition after interaction with both scatterers is $\Psi_{\textrm{scattered}} \propto \psi_{12}^{scattered}\psi_{3} + \psi_{13}^{scattered}\psi_{2}$.

The first term, corresponding to the particle scattering from $M_{2}$, is then a product of the initial $M_{3}$ scatterer substate, $\psi_{3}$, with the entangled scattered substate $\psi_{12}^{scattered}$, constrained by the three-body Schr\"odinger equation
\begin{equation}
\begin{gathered}
(\frac{\hbar \partial_{x_{1}}^{2}}{2m}+\frac{\hbar \partial_{x_{2}}^{2}}{2M_{2}}+ PE[x_{1}-x_{2}]
+i\partial_{t})\psi_{12}^{scattered}=0. \notag
\label{eq:Scheqn}
\end{gathered}
\end{equation}
For hard-sphere scattering $PE[x_{1}-x_{2}]$ is proportional to a delta function but is effectively a boundary condition requiring the sum of the appropriate incident and reflected free body substates to vanish at the scatterer.

The solution $\psi_{12}^{scattered}\psi_{3}$ is given by \cite{kowalski1,kowalski2}
\begin{equation} 
\psi_{3} \propto \exp[i (K_{3} x_{3}-\frac{\hbar K_{3}^{2}}{2M_{3}}t)] \nonumber,
\end{equation}
and
\begin{equation}
\psi_{12}^{scattered} \propto \exp[i (k_{{\bf r}} x_{1}-\frac{\hbar k_{{\bf r}}^{2}}{2m}t+K_{{\bf 2r}} x_{2}-\frac{\hbar K_{{\bf 2r}}^{2}}{2M_{2}}t)],  \nonumber
\label{eq:state1}
\end{equation}
where $ k_{{\bf r}}=m (2M_{2}V_{2}-M_{2}v+mv)/\hbar(M_{2}+m)$ and $ K_{{\bf 2r}}=M_{2}(M_{2}V_{2}-mV_{2}+2mv)/\hbar(M_{2}+m)$. These wavevectors are obtained by determining the velocities of the reflected particle and scatterer, $v_{{\bf r}}$ and $V_{{\bf 2r}}$, from conservation of momentum and energy in the elastic collision and then used in the relations $ k_{{\bf r}}=mv_{{\bf r}}/\hbar$ and $ K_{{\bf 2r}}=M_{2}V_{{\bf 2r}}/\hbar$. The result after the particle scatters from $M_{3}$, given by $\psi_{13}^{scattered}\psi_{2}$, is similarly obtained. 

Gaussian wavegroup substates, constructed from these eigenstates of momentum, then provide constraints on the phase in $\psi_{12}^{scattered}$ and $\psi_{13}^{scattered}$: the wavegroup peaks must reflect as classical particles. In addition, the boundary condition must be satisfied. 

Both conditions require introducing offsets in the incident and reflected three-body wavefunctions. For example, the incident state, $\psi_{1}[x_{1}]\psi_{2}[x_{2}] \psi_{3}[x_{3}]$, requires the substitution $x_{3}\rightarrow x_{3}-x_{0}$ for the second scatterer wavegroup to be offset from the origin at $t=0$. Similarly, the scattered state $\psi_{12}^{scattered}[x_{1},x_{2}]\psi_{3}[x_{3}]$ requires $x_{3}\rightarrow x_{3}-x_{0}$. However, since at $t=0$ the particle wavegroup interacts with the first scatterer at the origin there are no offsets in $x_{1}$ or $x_{2}$. 

For $\psi_{13}^{scattered}[x_{1},x_{3}]\psi_{2}[x_{2}]$ the non-interacting scatterer wavegroup, initially located at the origin, requires no offset.  However, the delay in the particle wavegroup reaching the scatterer wavegroup which is located at $x_{0}$ when $t=0$ requires $x_{3}\rightarrow x_{3}-x_{30}$ and $x_{1}\rightarrow x_{1}-x_{10}$. These offsets are determined from classical reflected particle and scatterer position functions, $x_{10}+v_{{\bf 1r}}t$, $x_{30}+V_{{\bf 3r}}t$ and constraining them to be equal to classical position functions of the incident particle and second scatterer at the time of interaction, resulting in $x_{10}=2Mx_{0}/(m+M)$ and $x_{30}=(M-m)x_{0}/(m+M)$.

The interference of these two three-body scattered momentum eigenstates is then given by
\begin{equation}
\begin{gathered}
\textrm{PDF}[x_{1},x_{2},x_{3}] \propto \cos^{2}[A/B],\label{eq:1} \\
A=\Delta V_{12}  \Delta x_{12} m^{2}M_{2}-\Delta V_{13} \Delta x_{13} m^{2}M_{3}+ \\ mM_{2}M_{3}(V_{3} x_{1}-v x_{2}-V_{2} \Delta x_{12} + \Delta V_{13}x_{3}), \\
B =\hbar(m+M_{2})(m+M_{3}),
\end{gathered}
\end{equation}
where $\Delta V_{12}=v-V_{2}$, $\Delta x_{12}=x_{1}-x_{2}$, $\Delta V_{13}=v-V_{3}$, and $\Delta x_{13}=x_{1}-x_{3}$.

\section{\label{appendix:wavegoups}{\bf Three-body wavegroups}}

Coordinate space wavegroups for the two reflected states are constructed from a superposition of these eigenstates of momentum \cite{kowalski1}. To apply approximations such as $m \ll M_{2}$ or $ M_{3}$ this superposition is expressed in terms of the particle and scatterer velocities rather than their wavevectors,
\begin{equation}
\begin{gathered}
\scalebox{.95}{$
\Psi_{scattered}^{group} \propto \iiint (\psi_{12}^{scattered}\psi_{3} + \psi_{13}^{scattered}\psi_{2}) \exp[-\frac{(v-v_{0})^{2}}{2\Delta v^{2}}]$} \\
\scalebox{.95}{$\times \exp[-\frac{(V_{2}-V_{20})^{2}}{2\Delta V_{2}^{2}}]\exp[-\frac{(V_{3}-V_{30})^{2}}{2\Delta V_{3}^{2}}] dv dV_{2}dV_{3}$} \nonumber,
\end{gathered}
\end{equation}
where $v_{0}$, $V_{20}$, and $V_{30}$ are the peaks of the Gaussian velocity distributions and $\Delta v$, $\Delta V_{2}$ and $\Delta V_{3}$ are their widths. The PDF, assuming $M_{2}$ is a rigid classical potential acting as a beamsplitter and resting at the origin, for $t=0$ and $V_{30}=0$ simplifies to
\begin{equation}
\begin{gathered}
\scalebox{.9}{$\textrm{PDF}\propto$}~\scalebox{1.07}{$
\exp[-A]+\exp[-B]+\exp[-C]\cos[2mv_{0}x_{3}/\hbar]$},\\
\scalebox{.95}{$A=2\Delta V_{3}^{2}M_{3}^{2}(x-x_{0})^{2}/\hbar^{2}$},\\
\scalebox{.95}{$B=2[(2 m x_{3}\Delta v)^{2}+\Delta V_{3}^{2}(M_{3}x_{0}+2mx_{3}-M_{3}x_{3})^{2}]/\hbar^{2}$}, \\
\scalebox{.95}{$C=2[2(m x_{3}\Delta v)^{2}+\Delta V_{3}^{2}(M_{3}^{2}(x_{0}-x_{3})^{2}]/\hbar^{2}$}\\ \scalebox{.95}{$+2mM_{3}(x_{0}-x_{3})x_{3}+2m^{2}x_{3}^{2}$}.\nonumber
\end{gathered}
\end{equation} \\
Eqn. (\ref{eqn:correlation2}) then follows from the approximation $m\ll M_{3}$ and where use is made of $m v=2\pi \hbar/\lambda_{0}$.

\begin{acknowledgments}
I wish to acknowledge many thoughtful comments from Nicholas Materise in reviewing the manuscript. 
\end{acknowledgments}

\end{document}